\documentclass[twocolumn,10pt]{article}

\pdfoutput=1

\usepackage[T1]{fontenc}
\usepackage[utf8]{inputenc}
\usepackage[a4paper,margin=1.65cm]{geometry}
\usepackage{amsmath,amssymb,bm}
\usepackage{graphicx}
\usepackage{booktabs}
\usepackage{placeins}
\usepackage[numbers,sort&compress]{natbib}
\usepackage[hidelinks]{hyperref}
\newcommand{\doilink}[1]{\href{https://doi.org/#1}{\nolinkurl{doi:#1}}}
\allowdisplaybreaks
\graphicspath{{./}}
\title{Finite mass two throat wormholes: global light rings, branch resolved strong lensing, and scalar transmission}
\author{ Anirudh Pradhan\textsuperscript{1}, K. Ghaderi\textsuperscript{2,*}, M. Zeyauddin\textsuperscript{3}, and  K. Karimizadeh\textsuperscript{4}}
\date{}

\begin{document}
\twocolumn[
\begin{@twocolumnfalse}
\maketitle
\vspace{-1.5em}
\begin{center}

\textsuperscript{1}Centre for Cosmology, Astrophysics and Space Science, GLA University, Mathura 281406, U. P., India\\
\textsuperscript{2}Department of Physics, Mari.C., Islamic Azad University, Marivan, Iran\\
\textsuperscript{3}Department of General Studies (Mathematics), Jubail Industrial College, Jubail 31961, Saudi Arabia\\
\textsuperscript{4} Department of Electrical Engineering,  Mari.C., Islamic Azad University, Marivan, Iran\\
\textsuperscript{*}Corresponding author: \href{mailto:k.ghaderi@iau.ac.ir}{k.ghaderi@iau.ac.ir}
\end{center}
\begin{abstract}
We introduce a static, horizonless, asymptotically flat two throat wormhole family in which one global metric determines the geometry, effective source, null dynamics, lensing, and test field propagation. A finite mass deformation of a quartic embedding profile produces two symmetric throats and an intermediate equator, while the associated Misner--Sharp mass tends to the same finite value at both asymptotic ends. The temporal sector yields the same mass parameter, and the Einstein tensor defines a single conserved anisotropic source. In addition to the local flare out identity, the complete radial averaged null functional is obtained exactly and is strictly negative. The global null potential admits a complete phase classification for all positive masses and nonnegative redshift deformation. Depending on the parameters, the global unstable set consists of throat rings, four off throat rings, or exterior rings accompanied by a subcritical equatorial ring. In the four ring phase, all unstable rings have the same impact scale but the inner and outer branches have different Lyapunov exponents. We derive the logarithmic strong deflection coefficients analytically and show that the cross throat coefficient is the sum of the contributions from every global maximum traversed by the ray. Direct numerical integration verifies these coefficients. The scalar scattering problem exhibits phase dependent barriers and resonant transmission. The resulting model provides a finite mass setting in which topology, averaged energy conditions, branch resolved strong lensing, and wave transmission are derived from one spacetime.
\end{abstract}
\vspace{0.5em}
\vspace{1.5em}
\end{@twocolumnfalse}
]
\section{Introduction}
\label{sec:introduction}

Traversable wormholes provide a controlled setting in which nontrivial spatial topology, energy conditions, and causal propagation can be examined without an event horizon. In a static spherical geometry, a throat is a local minimum of the areal radius. The flare out condition then implies radial null energy condition violation in general relativity \cite{Morris1988,Morris1988PRL,HochbergVisser1998,Visser1995}. The local theorem does not determine the integrated violation, which depends on the complete geometry and the affine weighting of the null generator \cite{VisserKarDadhich2003,Lobo2008}.

Most analytic models contain one throat. Two throat geometries are richer because the areal radius has two minima separated by an equator. Embedding constructions make this structure transparent: a prescribed areal profile fixes all stationary surfaces before the source is read from the field equations \cite{Rueda2022,Cataldo2026}. The quartic profile discussed in Ref.~\cite{Cataldo2026} is asymptotically flat in the sense that $b/r\rightarrow0$, but its shape function tends to zero. A finite mass extension is therefore needed if the geometry is to be treated as a compact lens with an independently defined asymptotic mass.

The optical analysis must also be global. Circular null orbits are extrema of the null potential with respect to a regular radial coordinate. Consequently, every throat and equator is a candidate light ring, in addition to roots obtained on monotonic areal branches \cite{Tsukamoto2024,Xavier2024}. Wormholes and other horizonless ultracompact objects can contain multiple photon and antiphoton spheres, producing several strong lensing families \cite{Shaikh2019PLB,Shaikh2019PRD,Tsukamoto2021,Guerrero2022,Huang2023}. A calculation restricted to one exterior branch can miss both the multiplicity and the dynamical inequivalence of the global rings.

Strong lensing near an ordinary unstable photon sphere has a logarithmic divergence governed by a local coefficient \cite{Bozza2002}. For a two ended wormhole, however, there are two geometrically different channels. A ray may turn around and return to the observer's asymptotic region, or it may cross the complete throat system and emerge at the opposite end. Same side and opposite side source configurations are established components of wormhole lensing \cite{Shaikh2019JCAP,Chen2025}. The critical curve alone does not specify an observable dark region because the intensity inside it depends on illumination and absorption in the second asymptotic region \cite{CunhaHerdeiro2018,PerlickTsupko2022}.

This work develops a finite mass two throat geometry and analyzes all of these sectors using one metric. The main additions beyond a basic geometric construction are fourfold. First, the local energy condition analysis is supplemented by an exact complete radial averaged null identity. Second, the light ring phase diagram is classified for the full domain $m>0$ and $\chi\ge0$, including the degenerate phase boundaries. Third, coordinate time Lyapunov exponents and logarithmic strong deflection coefficients are obtained for each branch. This reveals a phase in which four rings share one critical impact parameter but have different instability rates and different logarithmic weights. Fourth, the corresponding scalar scattering problem is solved as a two ended transmission problem and displays resonant structure associated with the multi barrier potential.

The paper is organized as follows. Section~\ref{sec:geometry} defines the global geometry, masses, and curvature diagnostics. Section~\ref{sec:source} derives the effective source and the exact radial averaged null identity. Section~\ref{sec:lightrings} gives the complete light ring classification and branch stability measures. Section~\ref{sec:lensing} develops reflected and transmitted strong lensing, including analytic logarithmic coefficients. Section~\ref{sec:scalar} treats scalar scattering. Section~\ref{sec:discussion} assesses the physical content and scope of the construction, and Sec.~\ref{sec:conclusion} summarizes the results.

\section{Global finite mass two throat geometry}
\label{sec:geometry}

We use $c=1$ and introduce a reference throat radius $r_0$. Dimensionless variables are
\begin{equation}
\begin{gathered}
 \bar t=\frac{t}{r_0},\quad
 \bar r=\frac{r}{r_0},\quad
 \bar z=\frac{z}{r_0},\\[2pt]
 \bar a=\frac{a}{r_0^2},
 \quad \bar K=K r_0^3,
 \quad m=\frac{M}{r_0}.
\end{gathered}
 \label{eq:dimensionless}
\end{equation}
The global line element is
\begin{align}
 \frac{ds^2}{r_0^2}&=-A(\bar r)d\bar t^2+C(\bar z)d\bar z^2
 +\bar r^2(\bar z)d\Omega^2,\notag\\
 C(\bar z)&=1+\bar r_{,\bar z}^{\,2}.
 \label{eq:globalmetric}
\end{align}
The coordinate $\bar z$ covers both asymptotic regions and remains regular at every extremum of the areal radius.

Define
\begin{equation}
 X=\bar z^2-\bar a,
 \qquad q=64\bar K^2m^2,
 \label{eq:Xq}
\end{equation}
and prescribe the smooth profile
\begin{equation}
 \bar r(\bar z)=1+\frac{\bar K X^2}{\sqrt{1+qX^2}},
 \qquad \bar a>0,\quad \bar K>0,\quad m>0.
 \label{eq:embedding}
\end{equation}
For fixed finite $\bar z$, Eq.~\eqref{eq:embedding} tends to the quartic embedding profile as $m\rightarrow0$. For nonzero $m$, its large $|\bar z|$ behavior changes from quartic to quadratic and generates a finite asymptotic mass.

The first derivative is
\begin{equation}
 \bar r_{,\bar z}=
 \frac{2\bar K\bar z X(2+qX^2)}{(1+qX^2)^{3/2}}.
 \label{eq:rz}
\end{equation}
There are three stationary surfaces:
\begin{align}
 \bar z_{\rm th}^{\pm}&=\pm\sqrt{\bar a},
 &\bar r(\bar z_{\rm th}^{\pm})&=1,
 \notag\\
 \bar r_{,\bar z\bar z}(\bar z_{\rm th}^{\pm})&=8\bar K\bar a>0,
 \label{eq:throats}\\[2pt]
 \bar z_*&=0,
 &\bar r_*&=1+\frac{\bar K\bar a^2}{\sqrt{1+q\bar a^2}},
 \notag\\
 \bar r_{,\bar z\bar z}(0)&<0,
 \notag\\[-2pt]
 \bar r_{,\bar z\bar z}(0)&=-\frac{2\bar K\bar a(2+q\bar a^2)}{(1+q\bar a^2)^{3/2}}.
 \label{eq:equator}
\end{align}
Thus, $\bar z=\pm\sqrt{\bar a}$ are throats and $\bar z=0$ is the intermediate equator.

The parametric Morris--Thorne shape function is
\begin{equation}
 \frac{\bar b(\bar z)}{\bar r(\bar z)}=\frac{1}{1+\bar r_{,\bar z}^{\,2}},
 \qquad
 \bar b(\bar z)=\frac{\bar r(\bar z)}{1+\bar r_{,\bar z}^{\,2}}.
 \label{eq:bparametric}
\end{equation}
The areal coordinate is not global because $\bar r$ is nonmonotonic. Equality $\bar b=\bar r$ occurs at the throats and at the equator; the sign of the second proper radial derivative distinguishes minima from maxima. With
\begin{equation}
 d\bar\ell=\sqrt{C}\,d\bar z,
 \label{eq:properdistance}
\end{equation}
the throat derivative is
\begin{equation}
 \left.\frac{d\bar b}{d\bar r}\right|_{\rm th}=1-16\bar K\bar a<1.
 \label{eq:bprime}
\end{equation}
The flare out condition therefore holds analytically for every $\bar K\bar a>0$.

At large $|\bar z|$,
\begin{align}
 \bar r(\bar z)&=\frac{\bar z^2}{8m}+1-\frac{\bar a}{8m}+O(\bar z^{-2}),
 \label{eq:rasymptotic}\\
 \bar b(\bar z)&=2m+O(\bar z^{-2}),
 \label{eq:basymptotic}\\
 g_{\bar r\bar r}&=1+\frac{2m}{\bar r}+O(\bar r^{-2}).
 \label{eq:grrasymptotic}
\end{align}
The Misner--Sharp mass is $M_{\rm MS}=r_0\bar b/2$, and hence tends to $M=mr_0$ at both ends.
This coefficient is also the ADM mass inferred from the spatial asymptotics in Eq.~\eqref{eq:grrasymptotic}.

We choose
\begin{equation}
 \Phi(\bar r)=-\frac{m}{\bar r}-\frac{\chi}{\bar r^2},
 \qquad A(\bar r)=e^{2\Phi(\bar r)},
 \qquad \chi\ge0.
 \label{eq:redshift}
\end{equation}
Since $\bar r\ge1$, the lapse is finite and positive. Its asymptotic expansion is
\begin{equation}
 A(\bar r)=1-\frac{2m}{\bar r}
 +\frac{2m^2-2\chi}{\bar r^2}+O(\bar r^{-3}),
 \label{eq:Aasymptotic}
\end{equation}
so the Komar mass at either infinity has the same value $M=mr_0$. The mass is therefore not introduced through an observational calibration.

Regularity follows directly from the global form. The functions $\bar r$, $A$, and $C$ are smooth, with $\bar r\ge1$, $A>0$, and $C\ge1$. In proper radial coordinates, define
\begin{equation}
\begin{gathered}
 \mathcal X=\Phi_{,\ell\ell}+\Phi_{,\ell}^2,
 \quad
 \mathcal Y=\frac{\Phi_{,\ell}\bar r_{,\ell}}{\bar r},\\[2pt]
 \mathcal Z=\frac{\bar r_{,\ell\ell}}{\bar r},
 \quad
 \mathcal W=\frac{1-\bar r_{,\ell}^2}{\bar r^2}.
\end{gathered}
 \label{eq:curvatureblocks}
\end{equation}
The dimensionless Kretschmann scalar is
\begin{equation}
 r_0^4 R_{\mu\nu\rho\sigma}R^{\mu\nu\rho\sigma}
 =4\left(\mathcal X^2+2\mathcal Y^2+2\mathcal Z^2+\mathcal W^2\right),
 \label{eq:Kretschmann}
\end{equation}
and is finite everywhere. The radial proper distance diverges at both asymptotic ends.

\begin{figure*}[t]
 \centering
 \includegraphics[width=0.96\textwidth]{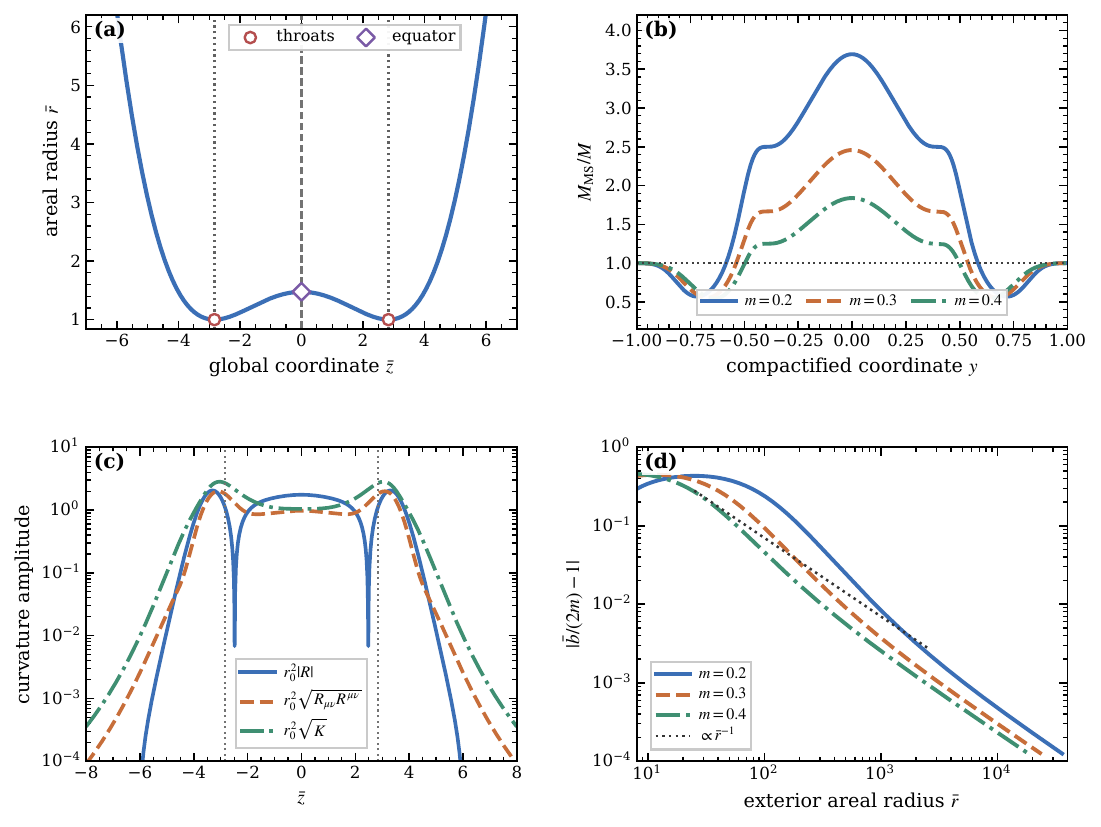}
 \caption{Geometry and regularity diagnostics. (a) Global areal radius for the benchmark \eqref{eq:benchmarkgeometry}; dotted lines mark the throats and the dashed line marks the equator. (b) Misner--Sharp mass normalized by its asymptotic value for three masses. (c) Ricci, Ricci tensor, and Kretschmann amplitudes for $\chi=0.45$. (d) Convergence of the shape function to $2m$ on an exterior branch.  All curvature diagnostics remain finite at the stationary surfaces and decay at infinity.}
 \label{fig:geometry}
\end{figure*}

Figure~\ref{fig:geometry} displays the geometry, mass function, curvature amplitudes, and asymptotic convergence. The benchmark used throughout is
\begin{equation}
 \bar a=8,\qquad \bar K=7.5\times10^{-3},\qquad m=0.3,
 \label{eq:benchmarkgeometry}
\end{equation}
for which $\bar r_*=1.4750994431$.

\section{Effective source and averaged null condition}
\label{sec:source}

The effective source is defined by the Einstein tensor of Eq.~\eqref{eq:globalmetric}. In an orthonormal frame,
\begin{equation}
 T^{\hat\mu}{}_{\hat\nu}=\mathrm{diag}(-\rho,p_r,p_t,p_t).
 \label{eq:Tdiag}
\end{equation}
No separately prescribed static dust sector is appended. Conservation follows identically from the contracted Bianchi identities.

Proper radial derivatives satisfy
\begin{equation}
 \bar r_{,\ell}=\frac{\bar r_{,\bar z}}{\sqrt C},
 \qquad
 \bar r_{,\ell\ell}=\frac{\bar r_{,\bar z\bar z}}{C^2},
 \label{eq:rproperderivatives}
\end{equation}
and
\begin{equation}
 \Phi_{,\ell}=\Phi_{,\bar r}\bar r_{,\ell},
 \qquad
 \Phi_{,\ell\ell}=\Phi_{,\bar r\bar r}\bar r_{,\ell}^2
 +\Phi_{,\bar r}\bar r_{,\ell\ell}.
 \label{eq:phiproper}
\end{equation}
The independent Einstein equations are
\begin{align}
 8\pi G r_0^2\rho&=
 \frac{1-\bar r_{,\ell}^2-2\bar r\bar r_{,\ell\ell}}{\bar r^2},
 \label{eq:rho}\\
 8\pi G r_0^2p_r&=
 \frac{\bar r_{,\ell}^2-1}{\bar r^2}
 +\frac{2\Phi_{,\ell}\bar r_{,\ell}}{\bar r},
 \label{eq:pr}\\
 8\pi G r_0^2p_t&=
 \Phi_{,\ell\ell}+\Phi_{,\ell}^2
 +\frac{\Phi_{,\ell}\bar r_{,\ell}}{\bar r}
 +\frac{\bar r_{,\ell\ell}}{\bar r}.
 \label{eq:pt}
\end{align}
These quantities are finite at all stationary surfaces.

At a stationary surface of $\bar r$,
\begin{equation}
 8\pi G r_0^2(\rho+p_r)
 =-\frac{2\bar r_{,\ell\ell}}{\bar r}.
 \label{eq:radialNECstationary}
\end{equation}
At either throat,
\begin{align}
 8\pi G r_0^2\rho_{\rm th}&=1-16\bar K\bar a,
 \label{eq:rhoth}\\
 8\pi G r_0^2p_{r,\rm th}&=-1,
 \label{eq:prth}\\
 8\pi G r_0^2p_{t,\rm th}&=8\bar K\bar a(1+m+2\chi),
 \label{eq:ptth}\\
 8\pi G r_0^2(\rho+p_r)_{\rm th}&=-16\bar K\bar a<0,
 \label{eq:NECth}\\
 8\pi G r_0^2(\rho+p_t)_{\rm th}&=1+8\bar K\bar a(m+2\chi-1),
 \label{eq:tangentialNECth}\\
 8\pi G r_0^2(\rho+p_r+2p_t)_{\rm th}&=16\bar K\bar a(m+2\chi).
 \label{eq:SECth}
\end{align}
For the benchmark, $d\bar b/d\bar r|_{\rm th}=0.04$, $8\pi G r_0^2\rho_{\rm th}=0.04$, and $8\pi G r_0^2(\rho+p_r)_{\rm th}=-0.96$.

A complete radial null geodesic supplies a stronger global statement. Up to a positive normalization of the affine parameter, define the dimensionless averaged null functional
\begin{equation}
 \mathcal A=\int_{-\infty}^{\infty}
 e^{-\Phi}\left[8\pi G r_0^2(\rho+p_r)\right]d\bar\ell.
 \label{eq:ANECdef}
\end{equation}
Using Eqs.~\eqref{eq:rho} and \eqref{eq:pr},
\begin{equation}
 8\pi G r_0^2(\rho+p_r)=
 2\left(\frac{\Phi_{,\ell}\bar r_{,\ell}}{\bar r}
 -\frac{\bar r_{,\ell\ell}}{\bar r}\right).
 \label{eq:radialNECgeneral}
\end{equation}
An integration by parts gives
\begin{equation}
 \mathcal A=-2\int_{-\infty}^{\infty}
 e^{-\Phi}\left(\frac{\bar r_{,\ell}}{\bar r}\right)^2d\bar\ell<0.
 \label{eq:ANECidentity}
\end{equation}
The boundary term vanishes as $1/\bar r$ at each end. Equation~\eqref{eq:ANECidentity} is exact for the full family and shows that positive values of $\rho+p_r$ near the equator cannot reverse the complete radial average.

\begin{figure*}[t]
 \centering
 \includegraphics[width=0.96\textwidth]{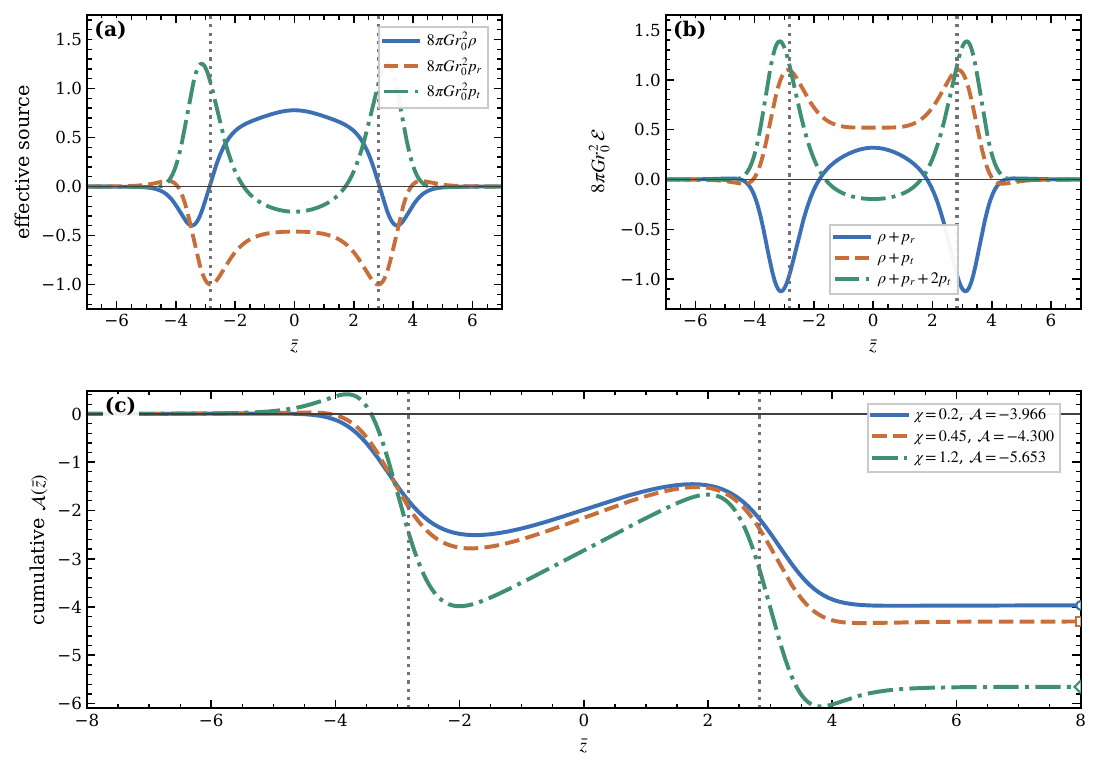}
 \caption{Effective source and averaged null diagnostics for the benchmark geometry. (a) Density and principal pressures at $\chi=0.45$. (b) Null and strong energy condition combinations. (c) Cumulative radial averaged null functional for three redshift deformations; the asymptotic values are reported in the legend. Dotted lines mark the throats, and line styles provide a color independent distinction.}
 \label{fig:source}
\end{figure*}

Figure~\ref{fig:source} displays the source and cumulative integral. For $\chi=0.2$, $0.45$, and $1.2$, respectively, the complete values are $\mathcal A=-3.966138$, $-4.299952$, and $-5.653192$.

\section{Complete light ring phase structure}
\label{sec:lightrings}

For equatorial null geodesics, define
\begin{equation}
 U(\bar z)=\frac{A[\bar r(\bar z)]}{\bar r^2(\bar z)},
 \qquad
 \mathcal C(\bar r)=\bar r\Phi_{,\bar r}
 =\frac{m}{\bar r}+\frac{2\chi}{\bar r^2}.
 \label{eq:Ucompactness}
\end{equation}
The derivative of the potential is
\begin{equation}
 U_{,\bar z}=\frac{2U\bar r_{,\bar z}}{\bar r}\left(\mathcal C-1\right).
 \label{eq:Uprime}
\end{equation}
Circular null orbits therefore occur either at a stationary surface of $\bar r$ or at
\begin{equation}
 \mathcal C(\bar r_{\rm ph})=1,
 \qquad
 \bar r_{\rm ph}=\frac{m+\sqrt{m^2+8\chi}}{2}.
 \label{eq:rph}
\end{equation}
Since $\mathcal C_{,\bar r}<0$, every off throat solution of Eq.~\eqref{eq:rph} is unstable. At a stationary surface $\bar z_s$,
\begin{equation}
 U_{,\bar z\bar z}(\bar z_s)=
 \frac{2U_s\bar r_{,\bar z\bar z}(\bar z_s)}{\bar r_s}
 \left(\mathcal C_s-1\right).
 \label{eq:Usecondstationary}
\end{equation}
The throat is unstable when $m+2\chi<1$ and stable when $m+2\chi>1$. The equator is unstable when $\mathcal C(\bar r_*)>1$ and stable when $\mathcal C(\bar r_*)<1$.

Introduce
\begin{equation}
 \chi_{\rm th}(m)=\frac{1-m}{2},
 \qquad
 \chi_*(m)=\frac{\bar r_*^2-m\bar r_*}{2}.
 \label{eq:phaseboundaries}
\end{equation}
The complete classification for $m>0$ and $\chi\ge0$ is:
\begin{enumerate}
 \item For $0<m<1$ and $0\le\chi<\chi_{\rm th}$, the two throats are the global unstable light rings. The equator is stable and Eq.~\eqref{eq:rph} has no root in the spacetime. This is phase I.
 \item Phase II occurs in either of the two domains
 \begin{equation}
 \begin{split}
  &0<m\le1,\qquad \chi_{\rm th}<\chi<\chi_*,\\
  &1<m<\bar r_*(m),\qquad 0\le\chi<\chi_*.
 \end{split}
 \label{eq:phaseIIregions}
 \end{equation}
 In both cases, $1<\bar r_{\rm ph}<\bar r_*$ occurs on four branches. There are two inner and two outer unstable light rings with the same $U$ and the same critical impact parameter. The throats and equator are stable.
 \item Phase III occurs for $m<\bar r_*(m)$ and $\chi>\chi_*$, for $m=\bar r_*(m)$ and $\chi>0$, or for $m>\bar r_*(m)$ and every $\chi\ge0$. Only two off throat global maxima remain, one in each exterior. The equator is also unstable but has a lower potential than the exterior pair.
\end{enumerate}
For the benchmark embedding parameters, the equation $m=\bar r_*(m)$ gives $m=1.3985308$. Thus, the full phase map includes a phase II strip for $1<m<1.3985308$, whereas phase I is restricted to $m<1$.

The boundaries are degenerate but regular. At $\chi=\chi_{\rm th}$, each off throat pair merges with a throat and the local maximum is quartic. At $\chi=\chi_*$, the two inner off throat rings merge at the equator, again producing a quartic maximum, while the two exterior rings remain nondegenerate. The point $m=\bar r_*(m)$, $\chi=0$ is the endpoint of the equatorial merger line. The logarithmic formulas derived below apply to nondegenerate rings away from these boundaries. At a quartically degenerate maximum, the leading strong deflection divergence is nonlogarithmic and requires the corresponding marginal expansion \cite{Tsukamoto2020Marginal,Igata2026Degenerate}.

The critical impact parameter and coordinate angular frequency are
\begin{equation}
 \frac{u_i}{r_0}=\frac{1}{\sqrt{U_i}},
 \qquad
 \Omega_i r_0=\sqrt{U_i}.
 \label{eq:impactOmega}
\end{equation}
For a nondegenerate unstable ring, the coordinate time Lyapunov exponent follows from the quadratic expansion of the radial equation:
\begin{equation}
 \lambda_i r_0=
 \sqrt{-\frac{A_iU_{,\bar z\bar z,i}}{2C_iU_i}}.
 \label{eq:Lyapunov}
\end{equation}
This is the local instability rate relevant to the eikonal correspondence \cite{Cardoso2009}. Stable light rings also identify trapped perturbative channels associated with long lived response in horizonless ultracompact geometries \cite{CunhaBerti2017,Churilova2020}.

A central result of phase II is that equal impact scales do not imply equal dynamics. The inner and outer rings have the same $\bar r_{\rm ph}$ and $\Omega$, but $C$ and $U_{,\bar z\bar z}$ differ between branches. Their Lyapunov exponents are therefore distinct. Table~\ref{tab:ringdata} gives representative values.

\begin{table}[t]
\centering
\caption{Unstable light ring data for the benchmark geometry. Only inequivalent nonnegative $\bar z$ branches are listed. The local logarithmic weight is $\bar a_i=\Omega_i/\lambda_i$. The phase III equatorial ring is included although it is not the global maximum.}
\label{tab:ringdata}
\scriptsize
\setlength{\tabcolsep}{2.8pt}
\renewcommand{\arraystretch}{1.12}
\begin{tabular}{@{}cclcccc@{}}
\toprule
$\chi$ & ph. & branch & $\bar z_i$ & $\Omega_i r_0$ & $\lambda_i r_0$ & $\bar a_i$\\
\midrule
0.20 & I & throat & 2.828427 & 0.606531 & 0.230162 & 2.635231\\
0.45 & II & inner & 2.039010 & 0.477181 & 0.142953 & 3.338021\\
0.45 & II & outer & 3.441285 & 0.477181 & 0.230468 & 2.070489\\
1.20 & III & equator & 0 & 0.318672 & 0.103880 & 3.067707\\
1.20 & III & outer & 4.216583 & 0.325527 & 0.337766 & 0.963764\\
\bottomrule
\end{tabular}
\end{table}

\begin{figure*}[t]
 \centering
 \includegraphics[width=0.96\textwidth]{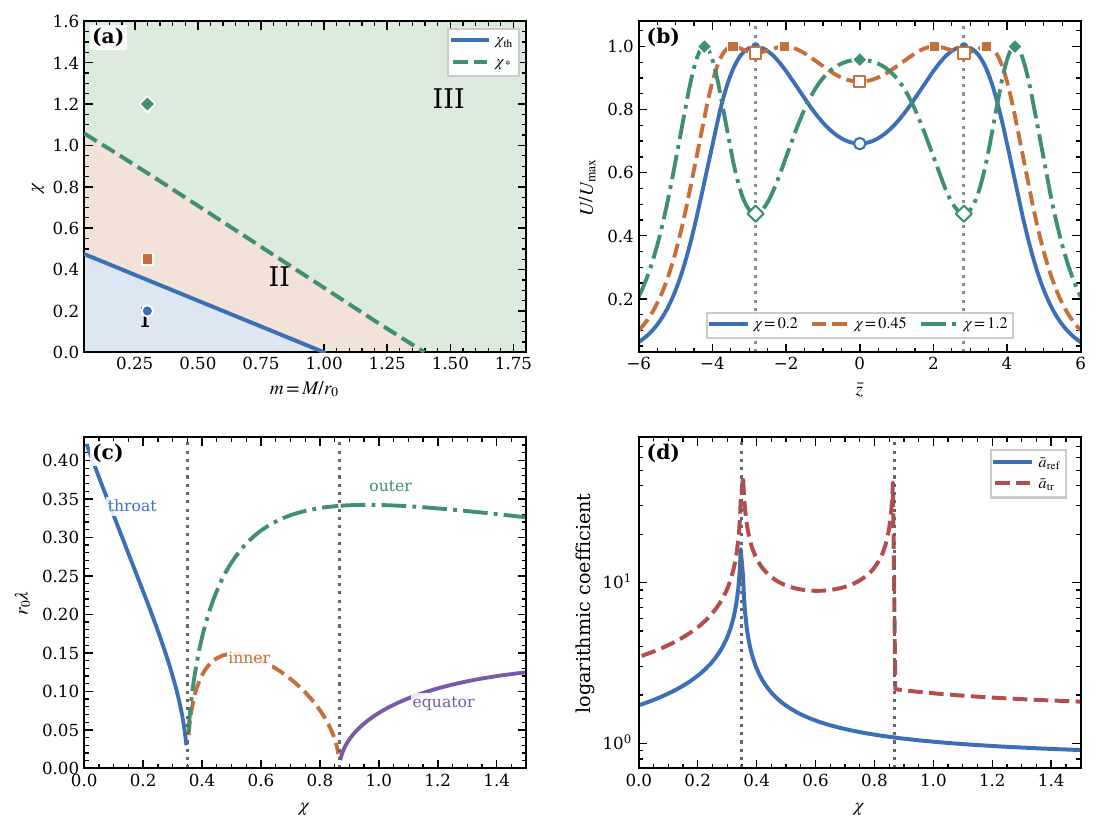}
 \caption{Global light ring phase structure for $\bar a=8$ and $\bar K=7.5\times10^{-3}$. (a) Complete phase map in the $(m,\chi)$ plane. (b) Normalized null potentials for the three benchmark phases; filled and open markers denote unstable and stable rings. (c) Coordinate time Lyapunov exponents of the inequivalent unstable branches at $m=0.3$. (d) Reflected and transmitted logarithmic strong deflection coefficients. Vertical dotted lines mark $\chi_{\rm th}$ and $\chi_*$.}
 \label{fig:lightrings}
\end{figure*}

Figure~\ref{fig:lightrings} shows the full phase map, the three representative potentials, the branch Lyapunov exponents, and the strong deflection coefficients derived below.

\section{Reflected and transmitted strong lensing}
\label{sec:lensing}

Let $\bar u=u/r_0=L/(Er_0)$. The azimuthal equation is
\begin{equation}
 \frac{d\varphi}{d\bar z}=
 \frac{\bar u\sqrt{AC}}
 {\bar r^2\sqrt{1-\bar u^2U}}.
 \label{eq:dphidz}
\end{equation}
For a ray that turns at $\bar z_{\min}$ and returns to the same asymptotic region,
\begin{equation}
 \hat\alpha(\bar u)=
 2\int_{\bar z_{\min}}^{\infty}\frac{d\varphi}{d\bar z}d\bar z-\pi,
 \qquad
 \bar u^2U(\bar z_{\min})=1.
 \label{eq:sameside}
\end{equation}
A ray transmitted from one end to the other has
\begin{equation}
 \Delta\varphi_{+-}(\bar u)=
 \int_{-\infty}^{\infty}\frac{d\varphi}{d\bar z}d\bar z.
 \label{eq:crossside}
\end{equation}
The global critical value is fixed by the global maximum $U_c$,
\begin{equation}
 \bar u_c=U_c^{-1/2}.
 \label{eq:uc}
\end{equation}
Rays with $\bar u>\bar u_c$ are reflected, while rays with $\bar u<\bar u_c$ cross the complete geometry.

Near a nondegenerate global maximum $\bar z_i$, define
\begin{equation}
 \bar a_i=
 \frac{\sqrt{2A_iC_i}}
 {\bar r_i^2\sqrt{-U_{,\bar z\bar z,i}}}
 =\frac{\Omega_i}{\lambda_i}.
 \label{eq:localstrongcoefficient}
\end{equation}
The last equality follows from Eqs.~\eqref{eq:impactOmega} and \eqref{eq:Lyapunov}; it is also consistent with invariant formulations of the logarithmic rate in terms of the local geometry and source at the orbit \cite{Igata2026Invariant}. For the reflected channel, set
\begin{equation}
 \epsilon_+=\frac{\bar u^2}{\bar u_c^2}-1.
 \label{eq:epsplus}
\end{equation}
The turning point approaches the outermost encountered global maximum and
\begin{equation}
 \hat\alpha=-\bar a_{\rm ref}\ln\epsilon_++\bar b_{\rm ref}+O(\epsilon_+\ln\epsilon_+).
 \label{eq:strongref}
\end{equation}
For the transmitted channel, define
\begin{equation}
 \epsilon_-=1-\frac{\bar u^2}{\bar u_c^2},
 \label{eq:epsminus}
\end{equation}
and write the winding number $N=\Delta\varphi_{+-}/(2\pi)$ as
\begin{equation}
 N=-\frac{\bar a_{\rm tr}}{2\pi}\ln\epsilon_-+\bar b_{\rm tr}
 +O(\epsilon_-\ln\epsilon_-).
 \label{eq:strongtrans}
\end{equation}
Each global maximum crossed by the ray contributes its local coefficient. Reflection samples one outer branch, whereas transmission samples all global maxima along the complete path. The phase dependent coefficients are therefore
\begin{align}
 \text{phase I:}\quad &\bar a_{\rm ref}=\bar a_{\rm th},
 &\bar a_{\rm tr}=2\bar a_{\rm th},
 \label{eq:coeffI}\\
 \text{phase II:}\quad &\bar a_{\rm ref}=\bar a_{\rm out},
 &\bar a_{\rm tr}=2(\bar a_{\rm in}+\bar a_{\rm out}),
 \label{eq:coeffII}\\
 \text{phase III:}\quad &\bar a_{\rm ref}=\bar a_{\rm out},
 &\bar a_{\rm tr}=2\bar a_{\rm out}.
 \label{eq:coeffIII}
\end{align}
Equation~\eqref{eq:coeffII} is particularly diagnostic: four rings are degenerate in $u_c$, but the transmitted divergence resolves their unequal local instabilities through the sum of inner and outer weights.

For the benchmark phases, Table~\ref{tab:lensingdata} lists the global impact scale and coefficients. The phase II transmitted coefficient is enhanced by more than a factor of five relative to the same side coefficient because the ray crosses four critical branches.

\begin{table}[t]
\centering
\caption{Global critical scale, strong deflection coefficients, and complete radial averaged null functional for the benchmark models.}
\label{tab:lensingdata}
\scriptsize
\setlength{\tabcolsep}{3.5pt}
\renewcommand{\arraystretch}{1.12}
\begin{tabular}{@{}cccccc@{}}
\toprule
$\chi$ & ph. & $\bar u_c$ & $\bar a_{\rm ref}$ & $\bar a_{\rm tr}$ & $\mathcal A$\\
\midrule
0.20 & I & 1.648721 & 2.635231 & 5.270463 & $-3.966138$\\
0.45 & II & 2.095643 & 2.070489 & 10.817021 & $-4.299952$\\
1.20 & III & 3.071944 & 0.963764 & 1.927528 & $-5.653192$\\
\bottomrule
\end{tabular}
\end{table}

\begin{figure*}[t]
 \centering
 \includegraphics[width=0.96\textwidth]{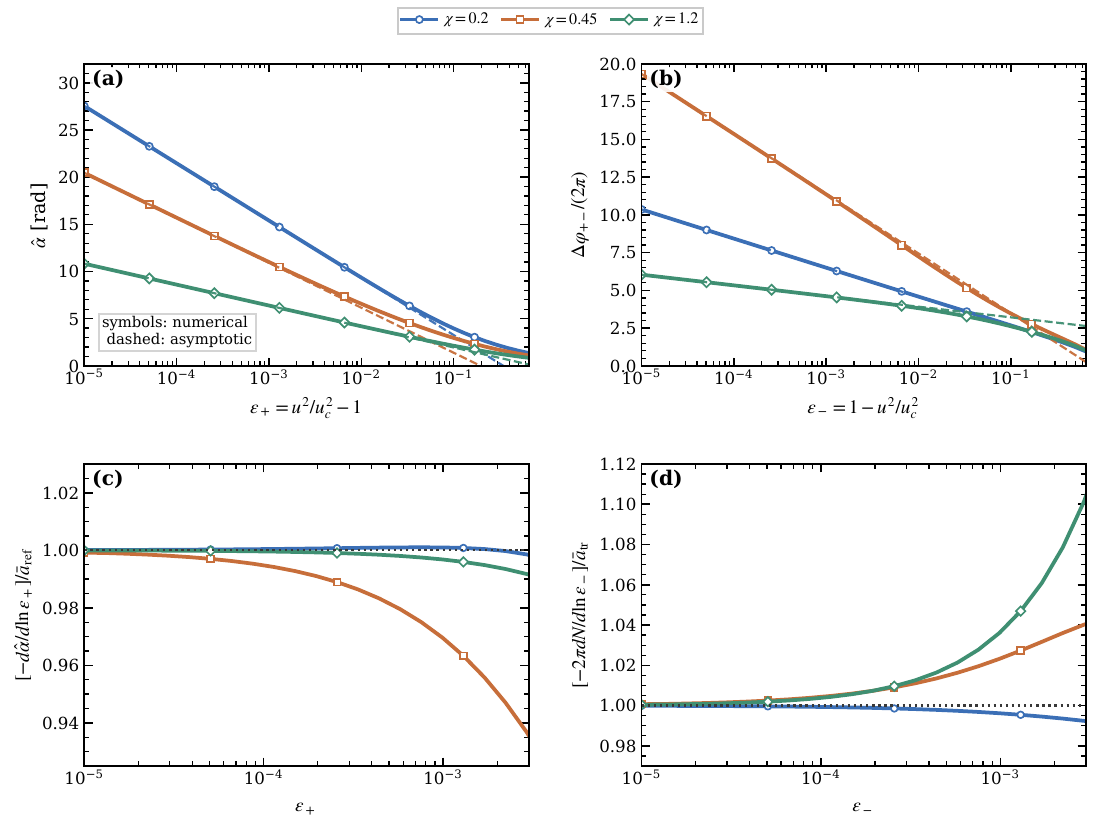}
 \caption{Strong lensing and asymptotic validation. (a) Same side deflection as the reflected critical branch is approached. (b) Cross throat winding number as the transmitted critical branch is approached. Dashed curves are the logarithmic asymptotics. (c) and (d) Numerical logarithmic slopes normalized by the analytic coefficients; unity is the predicted limit.}
 \label{fig:lensing}
\end{figure*}

Figure~\ref{fig:lensing} compares direct numerical integration with Eqs.~\eqref{eq:strongref} and \eqref{eq:strongtrans}. The logarithmic slopes converge to the analytic coefficients in all three phases.

\subsection{Critical curve and illumination}
\label{subsec:illumination}

For an observer in the positive asymptotic region, the critical angular radius is a property of the metric. The intensity assigned to the two sides is not. In the idealized case of two uniform celestial spheres with invariant source intensities $I_+$ and $I_-$, Liouville transport gives
\begin{equation}
 I_{\rm obs}(\bar u)=
 \begin{cases}
 I_-, & 0\le\bar u<\bar u_c,\\
 I_+, & \bar u>\bar u_c,
 \end{cases}
 \label{eq:twosky}
\end{equation}
up to redshift factors associated with finite source and observer positions. Thus the central domain is dark when the opposite region is dark or absorbing. Disk and hot spot images are then obtained by combining this global geodesic map with the chosen emission model and ray tracing prescription.

\section{Scalar transmission across the two ended geometry}
\label{sec:scalar}

Scalar transmission through asymptotically flat traversable wormholes can be formulated as a two ended scattering problem with reflection and transmission amplitudes \cite{Azad2020}. Resonant transmission caused by barrier supported trapping has also been reported in static and rotating wormhole backgrounds \cite{Karmakar2026}. Here we consider a minimally coupled massless scalar field, $\Box\Psi=0$, separated as
\begin{equation}
 \Psi=e^{-i\omega t}Y_{\ell m}(\theta,\varphi)\frac{\psi(x)}{r}.
 \label{eq:scalaransatz}
\end{equation}
The tortoise coordinate is defined by
\begin{equation}
 \frac{dx}{d\bar z}=r_0\sqrt{\frac{C}{A}}.
 \label{eq:tortoise}
\end{equation}
In dimensionless form, the radial equation is
\begin{equation}
 \frac{d^2\psi}{d(x/r_0)^2}+
 \left[(\omega r_0)^2-r_0^2V_\ell\right]\psi=0,
 \label{eq:scalarwave}
\end{equation}
with
\begin{equation}
 r_0^2V_\ell=A\left[
 \frac{\ell(\ell+1)}{\bar r^2}
 +\frac{1}{\bar r}\left(\bar r_{,\ell\ell}
 +\Phi_{,\ell}\bar r_{,\ell}\right)
 \right].
 \label{eq:scalarpotentialproper}
\end{equation}
Equivalently,
\begin{equation}
 r_0^2V_\ell=A\left[
 \frac{\ell(\ell+1)}{\bar r^2}
 +\frac{\Phi_{,\bar r}\bar r_{,\bar z}^2}{C\bar r}
 +\frac{\bar r_{,\bar z\bar z}}{C^2\bar r}
 \right].
 \label{eq:scalarpotentialz}
\end{equation}
The potential tends to zero at both ends. For incidence from the negative side,
\begin{align}
 \psi&\sim e^{i\omega x}+\mathcal R_\ell e^{-i\omega x},
 &&x\rightarrow-\infty,
 \label{eq:leftBC}\\
 \psi&\sim \mathcal T_\ell e^{i\omega x},
 &&x\rightarrow+\infty.
 \label{eq:rightBC}
\end{align}
The real potential gives $|\mathcal R_\ell|^2+|\mathcal T_\ell|^2=1$.

\begin{figure*}[t]
 \centering
 \includegraphics[width=0.96\textwidth]{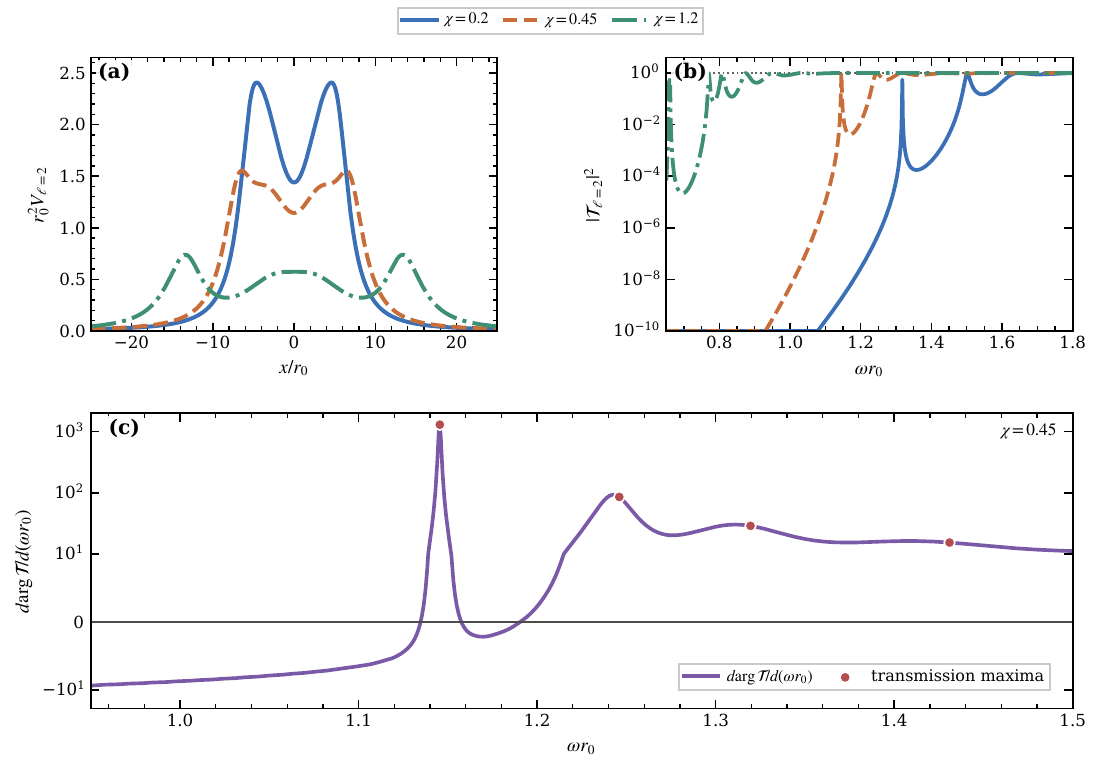}
 \caption{Scalar scattering for $\ell=2$. (a) Phase dependent effective potentials in the tortoise coordinate. (b) Transmission probabilities for the three benchmark models; the dotted line marks unit transmission. (c) Transmission phase derivative for phase II on a symmetric logarithmic vertical scale; markers identify local transmission maxima. The multi barrier geometry produces a sequence of resonant transmission features.}
 \label{fig:wave}
\end{figure*}

Figure~\ref{fig:wave} shows the $\ell=2$ potentials and transmission spectra. Phase I has a pronounced symmetric double barrier. Phase II retains a multi barrier structure and supports narrow transmission resonances. Phase III has lower exterior barriers and substantially larger broad band transmission. The phase derivative $d\arg\mathcal T/d(\omega r_0)$ resolves the phase II resonances. This scattering calculation provides a controlled test field diagnostic of the same global geometry.

\section{Discussion}
\label{sec:discussion}

The central consistency feature is that every sector is derived from one spacetime. Here the same functions $\bar r(\bar z)$ and $A(\bar r)$ determine the throat conditions, asymptotic masses, Einstein tensor, circular null orbits, lensing integrals, and scalar potential. There is no change of the radial metric between the source analysis and the optical calculation, and no pressureless static halo is introduced as a separately conserved component.

The construction is geometrical. The stress tensor is the exact conserved effective source required by the Einstein equations, and the metric, geodesic, source, and test field sectors are treated within this single effective spacetime. The exact identity \eqref{eq:ANECidentity} shows that local positive energy regions are accompanied by a definite complete radial averaged violation.

The new optical content is not the existence of opposite side wormhole lensing by itself. That channel is known in the literature. The distinctive result is the branch resolved structure of the present finite mass two throat geometry. In phase II, four global maxima have one critical impact parameter and one angular frequency, yet split into two Lyapunov classes. The same side divergence is controlled only by the outer class, whereas the transmitted divergence sums the inner and outer classes twice. This produces a large analytic enhancement of the cross throat logarithmic coefficient without introducing an independent optical prescription.

The scalar calculation provides an independent propagation diagnostic. The resonances arise from the multi barrier potential and are consistent with the stable trapping regions visible in the null potential. They are therefore presented as transmission properties of a minimally coupled scalar field on the finite mass two throat background.

The finite mass extension also resolves a scale ambiguity. Both asymptotic metric sectors identify $M=mr_0$, so ratios such as $u_c/M$ follow from the model parameters rather than from matching a desired angular diameter. The geometrical scale is consequently fixed internally by the model before any object specific application is attempted.

\section{Conclusions}
\label{sec:conclusion}

We constructed a smooth, horizonless, finite mass two throat wormhole in a global radial coordinate. The areal radius has two minima and one intermediate maximum, the Misner--Sharp mass tends to $M$ at both ends, and the temporal asymptotics yield the same Komar mass. Curvature invariants are finite throughout the spacetime.

The Einstein equations define one conserved anisotropic effective source. The local throat relation $8\pi G r_0^2(\rho+p_r)=-16\bar K\bar a$ follows directly from the flare out geometry. More globally, the complete radial averaged null functional has the exact negative form \eqref{eq:ANECidentity}.

The null potential was classified over the full $(m,\chi)$ domain. The three phases contain unstable throat rings, four off throat rings, or exterior rings together with an equatorial ring. The phase boundaries correspond to quartically degenerate mergers. In the four ring phase, equal impact parameters coexist with unequal inner and outer Lyapunov exponents.

The local instability rates determine analytic strong deflection coefficients through $\bar a_i=\Omega_i/\lambda_i$. Same side lensing samples one outer critical branch, while cross throat lensing sums every global maximum crossed by the ray. Numerical integrations reproduce the predicted logarithmic slopes. The same geometry also yields phase dependent scalar barriers and resonant transmission.

The result is a self contained finite mass model in which topology, source diagnostics, global light rings, strong lensing, and scalar propagation are mutually consistent. Its principal discriminant is the branch resolved dynamics of several light rings that can be degenerate in critical impact scale but inequivalent in instability and transmitted lensing weight.

\appendix

\section{Stationary surface and curvature identities}
\label{app:identities}

At any stationary surface $\bar z_s$, $\bar r_{,\bar z}=0$ and $C=1$. Differentiating Eq.~\eqref{eq:bparametric} gives
\begin{equation}
 \left.\frac{d\bar b}{d\bar r}\right|_{\bar z_s}
 =1-2\bar r_s\bar r_{,\bar z\bar z}(\bar z_s).
 \label{eq:bstationary}
\end{equation}
A minimum therefore obeys the flare out inequality, whereas a maximum has $d\bar b/d\bar r>1$ and is an equator.

The Ricci scalar can be written directly from the effective source,
\begin{equation}
 r_0^2R=8\pi G r_0^2(\rho-p_r-2p_t).
 \label{eq:RicciT}
\end{equation}
Writing $\mathcal E_\rho=8\pi G r_0^2\rho$ and analogously for the pressures,
\begin{multline}
 r_0^4R_{\mu\nu}R^{\mu\nu}=\frac14\big[
 (\mathcal E_\rho+\mathcal E_r+2\mathcal E_t)^2
 +(\mathcal E_\rho+\mathcal E_r-2\mathcal E_t)^2\\
 +2(\mathcal E_\rho-\mathcal E_r)^2\big].
 \label{eq:Ricci2}
\end{multline}
Equations~\eqref{eq:Kretschmann}, \eqref{eq:RicciT}, and \eqref{eq:Ricci2} provide independent regularity diagnostics.

\section{Lensing asymptotics and numerical control}
\label{app:lensingnumerics}

The same side integral has an integrable square root endpoint. We remove it with
\begin{equation}
 \bar z=\bar z_{\min}+s^2,
 \qquad d\bar z=2s\,ds.
 \label{eq:endpointsubstitution}
\end{equation}
At large positive $\bar z$,
\begin{equation}
 \frac{d\varphi}{d\bar z}=
 \frac{16m\bar u}{\bar z^3}+O(\bar z^{-5}),
 \label{eq:integrandtail}
\end{equation}
so the one sided tail beyond $Z$ is
\begin{equation}
 \int_Z^\infty\frac{d\varphi}{d\bar z}d\bar z
 =\frac{8m\bar u}{Z^2}+O(Z^{-4}).
 \label{eq:tail}
\end{equation}
For transmission, the integral is split at all stationary surfaces and the tail is added at both ends.

For the phase II benchmark, Table~\ref{tab:lensconvergence} shows cutoff convergence at $\bar u=0.9\bar u_c$ for transmission and $\bar u=1.1\bar u_c$ for reflection.

\begin{table}[htbp]
\centering
\caption{Cutoff convergence of the phase II lensing integrals. Absolute differences are measured relative to the $Z=260$ values.}
\label{tab:lensconvergence}
\scriptsize
\setlength{\tabcolsep}{2.6pt}
\renewcommand{\arraystretch}{1.12}
\begin{tabular}{@{}ccccc@{}}
\toprule
$Z$ & $\Delta\varphi_{+-}/(2\pi)$ & $|\Delta N|$ & $\hat\alpha$ & $|\Delta\hat\alpha|$\\
\midrule
80  & 2.605771627 & $2.04\times10^{-7}$ & 2.085568887 & $1.57\times10^{-6}$\\
120 & 2.605771793 & $3.79\times10^{-8}$ & 2.085570161 & $2.91\times10^{-7}$\\
180 & 2.605771825 & $5.98\times10^{-9}$ & 2.085570407 & $4.63\times10^{-8}$\\
260 & 2.605771831 & 0 & 2.085570452 & 0\\
\bottomrule
\end{tabular}
\end{table}
\FloatBarrier

To obtain Eq.~\eqref{eq:localstrongcoefficient}, expand
\begin{equation}
 U(\bar z)=U_i+\frac12U_{,\bar z\bar z,i}(\bar z-\bar z_i)^2+\cdots.
 \label{eq:Ulocalexpansion}
\end{equation}
The singular part of either a reflected passage or a complete crossing of one maximum is $-\bar a_i\ln\epsilon$. Summing the maxima encountered by each trajectory gives Eqs.~\eqref{eq:coeffI}--\eqref{eq:coeffIII}.

\section{Scalar transfer matrix and convergence}
\label{app:wavenumerics}

The potential is sampled on a uniform tortoise grid. Over a cell of width $h$ with constant $V_j$, the state vector $(\psi,d\psi/dx)^T$ is propagated by
\begin{equation}
 \bm P_j=
 \begin{pmatrix}
 \cos(q_jh) & \sin(q_jh)/q_j\\
 -q_j\sin(q_jh) & \cos(q_jh)
 \end{pmatrix},
 \qquad q_j^2=\omega^2-V_j.
 \label{eq:transfermatrix}
\end{equation}
The ordered product is matched to Eqs.~\eqref{eq:leftBC} and \eqref{eq:rightBC}. The determinant of each cell matrix is unity, and the numerical flux residual remains below $6\times10^{-12}$ over the displayed frequency range.

Table~\ref{tab:waveconvergence} reports the phase II transmission probability at $\omega r_0=1.145$ and $Z=24$. This test frequency lies close to the first narrow resonance and is therefore more sensitive to grid resolution than the broad band spectrum.

\begin{table}[htbp]
\centering
\caption{Grid convergence of $|\mathcal T_{\ell=2}|^2$ for the phase II benchmark at $\omega r_0=1.145$ and $Z=24$. The listed $h/r_0$ is an upper bound on the uniform grid spacing; the actual spacing differs by less than $1.4\times10^{-4}$ in relative terms. The last column is the relative deviation from the finest grid.}
\label{tab:waveconvergence}
\small
\setlength{\tabcolsep}{8.0pt}
\renewcommand{\arraystretch}{1.10}
\begin{tabular}{@{}ccc@{}}
\toprule
$h/r_0$ & $|\mathcal T_2|^2$ & relative deviation\\
\midrule
0.080 & 0.723404 & $1.84\times10^{-2}$\\
0.050 & 0.732092 & $6.60\times10^{-3}$\\
0.035 & 0.734940 & $2.73\times10^{-3}$\\
0.025 & 0.736281 & $9.13\times10^{-4}$\\
0.018 & 0.736954 & 0\\
\bottomrule
\end{tabular}
\end{table}
\FloatBarrier
The spectra in Fig.~\ref{fig:wave} use a uniform spacing not exceeding $h/r_0=0.018$ and $Z=24$. Panel (b) uses 1401 uniformly spaced frequencies over $0.65\le\omega r_0\le1.8$, while panel (c) uses 1101 frequencies over $0.95\le\omega r_0\le1.5$. Changing $Z$ from 20 to 30 at a spacing not exceeding $h/r_0=0.035$ changes the same test value by less than $6\times10^{-5}$.

\end{document}